\documentclass[epj,final]{svjour}
\usepackage{graphicx,amsmath,amssymb,bm}

\begin{document}
\bibliographystyle{revtex}

\title{Mechanical detection of nuclear spin relaxation in a
  micron-size crystal}

\author{O. Klein \inst{1} \thanks{\emph{Present address:} Service de
    Physique de l'Etat Condens\'e, CEA Orme des Merisiers, F-91191
    Gif-Sur-Yvette, \email{oklein@cea.fr}.}, V. V. Naletov \inst{2},
  and H.  Alloul\inst{1} \thanks{\emph{Present address:}Laboratoire de
    Physique des Solides, Universit\'e Paris-Sud, F-91405 Orsay.} }

\institute{Laboratoire des Solides Irradi\'es,
  Ecole Polytechnique, Palaiseau F-91128 \and Physics Department,
  Kazan State University, Kazan 420008 Russia}

\date{Submitted 22 February 2000}

\abstract{ A room temperature nuclear magnetic resonance force
  microscope (MRFM), fitted in a \( 1 \)Tesla electromagnet, is used
  to measure the nuclear spin relaxation of \( ^{1} \)H in a
  micron-size (70ng) crystal of ammonium sulfate. NMR sequences,
  combining both pulsed and continuous wave r.f. fields, have allowed
  us to measure mechanically $T_2$ and $T_1$, the transverse and
  longitudinal spin relaxation times. Because two spin species with
  different $T_1$ values are measured in our 7$\mu$m thick crystal,
  magnetic resonance imaging of their spatial distribution inside the
  sample section are performed. To understand quantitatively the
  measured signal, we carefully study the influence of the
  spin-lattice relaxation and the non-adiabaticity of the c.w.
  sequence on the intensity and time dependence of the detected
  signal.
   }

\maketitle

\section{Introduction}

For a long time research groups have looked for new ways of detecting
electronic or nuclear paramagnetic resonance with better sensitivity.
A review of different proposed methods can be found in the
introduction of Abragam's book \cite{abragam}. In two seminal papers
\cite{sidles91,sidles92}, Sidles recognized the advantage of coupling
the spin system to a mechanical oscillator for magnetic resonance
imaging. In this technique, the force signal is proportional to the
magnetic field gradient \cite{sidles93}, which, in extremely
inhomogeneous field, should allow high spatial resolution. The new
technique is referred to as magnetic resonance force microscopy (MRFM)
\cite{sidles95}.

The first magnetic resonance force signal was detected by Rugar
\textit{et al.} in 1992 while exciting electron spin resonance (eMRFM)
in a 30ng crystal of diphenylpicrylhydrazil \cite{rugar92}. Two years
later, Rugar \emph{et al.} reported the mechanical detection of
\(^{1}\)H (protons) nuclear magnetic resonance (nMRFM) in 12ng of
ammonium nitrate \cite{rugar94}. These two pioneering experiments
demonstrate that a micro-fabricated cantilever, identical to the ones
developed for atomic force microscopy, can detect the magnetic moment
of a micron-size sample. In the case of nuclear magnetic resonance
(NMR) \cite{rugar94}, the achieved sensitivity of $10^{13}$ spin, at
room temperature and in a field of 2.4T, represents a substantial
improvement over the standard coil detection.

%\subsection{Brief experimental review}

Significant progress was made in the past few years. In 1996, Zhang
\emph{et al.} mechanically detected the ferromagnetic resonance
(fMRFM) of yttrium iron garnet \cite{zhang96}. Imaging experiments
with eMRFM \cite{zuger93,hammel95}, nMRFM \cite{zuger96,schaff97} and
fMRFM \cite{suh98} were performed.  A magnetic resonance torque signal
in a homogeneous magnetic field \cite{ascoli96} was also detected.
Improvement of the force sensitivity by operating at low temperature
\cite{wago96,wago97,wago98} was demonstrated. Force maps of the sample
were obtained with the magnetic probe placed on the mechanical
resonator in eMRFM \cite{wago98b} and fMFRM \cite{suh98}. The highest
sensitivity reported so far is around 200 electron spin in a 1Hz
bandwidth. The result was obtained by operating an eMRFM at 77K in a
very large magnetic field gradient \cite{bruland98}. In 1996, Wago
\emph{et al.}  demonstrated that a pulse sequence combined with fast
adiabatic passages can allow to measure the nuclear spin-lattice
relaxation time of \( ^{19}\)F in calcium fluoride at low temperatures
\cite{wago96}. The same method was used to measure the longitudinal
spin relaxation of \( ^{1}\)H in ammonium sulfate at room temperature
and normal pressure \cite{schaff97b,verhagen99}.  Recent eMRFM work in
vitreous silica at 5K showed that the same principles can be also
applied to study electron spin dynamics of \( E^\prime \) centers with
long $T_1$ \cite{wago98}.

%\subsection{Present work}

In this paper, we report the first measurements of both the transverse
and longitudinal nuclear spin dynamics of \( ^{1}\)H using mechanical
detection. A very thin sample is used to analyze if new phenomena
might be specific to small sizes. Our instrument is a simple
home-built MRFM located inside a \(1\)Tesla electromagnet.  The
mechanical motion of the cantilever is monitored by a laser beam
deflection system. The sample is a 7$\mu$m thick crystal of
(NH\(_{4}\))\(_{2}\)SO\(_{4}\). Two spin-lattice relaxation times are
observed, \(T_{1s}=0.4\)s and \(T_{1l}=5\)s. The later value
corresponds to the $T_1$ reported in the literature for this compound
\cite{miller62,oreilly67}.  The short relaxation, however, might be
due to water contamination inside the crystal during its contact with
air.  These same two relaxation rates are also measured by
conventional NMR in powder samples with particles of dimensions
smaller than 50$\mu$m.

After introducing in Section \ref{section2} the measurement technique
employed in this study, we will present in Section \ref{section3} our
results on the transverse and longitudinal spin relaxation properties
of (NH\(_{4}\))\(_{2}\)SO\(_{4}\). This will be followed in Section
\ref{section4} by a more detailed analysis of the time dependence and
magnitude of the force signal in order to quantify the properties of
spin with short and long $T_1$ and to determine the effect of the
non-adiabaticity of the sequence in the measured signal. Finally a
model to describe our experimental data will be proposed.

\section{Measurement of the force signal} \label{section2}

%\subsection{Experimental setup}

\begin{figure}
\includegraphics*[width=8.3cm,draft=false]{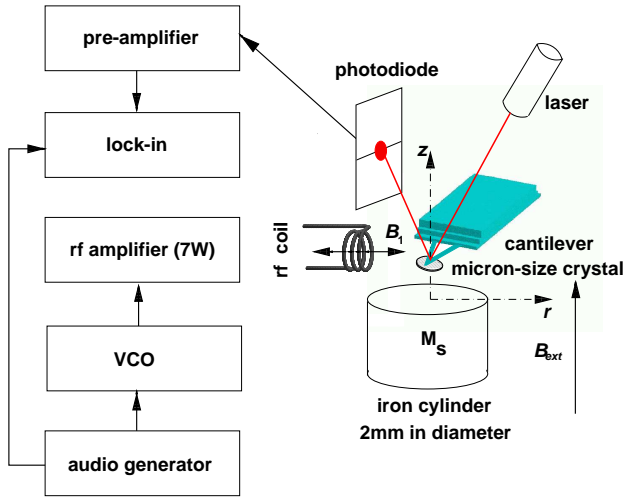}
\caption{MRFMs are miniature version of the Faraday balance. The instrument fits between the poles of  an electromagnet that generates  a homogeneous static field \(B_{ext}=\) 1T. A sample is affixed onto a micro-fabricated cantilever and placed in a field gradient $g=500$T/m produced by a 2 millimeter-size iron bar. The gradient serves to create a magnetic force on the spin system. The induced \AA-scale bending of the cantilever is measured by a laser beam deflection on a photodiode. To increase the sensitivity, the nuclear magnetization is inverted periodically at 1.4kHz, a frequency that corresponds to the fundamental flexure mode of the cantilever. The induced vibration is monitored by a lock-in. The nuclear spin oscillations are obtained by cyclic adiabatic inversions using frequency modulated r.f. field produced by a VCO. The carrier is at the Larmor frequency of protons.}
\label{FIG1}
\end{figure}

The setup is schematically represented in Fig.\ref{FIG1}. The
experiment is performed at room temperature inside a vacuum cell (10\(
^{-2} \)torr) constantly connected to the inlet of a rotary pump. The
instrument \cite{guillous99} fits between the poles of an iron core
electromagnet which produces a static magnetic field \( B_{ext}
\bm{k}\) along the $z$ axis. To the uniform field one adds a second
inhomogeneous field with axial symmetry produced by a magnetized iron
bar \( 8\)mm in length and \( 1.9\)mm in diameter.  The polarization
field experienced by the spin is \( B_0 = B_{ext} + B_{cyl} \), with
\( B_0 = \bm{B}_{0} \cdot \bm{k} \) and \( B_{cyl} = \bm{B}_{cyl}
\cdot \bm{k} \). Near the symmetry axis, the instantaneous magnetic
force acting on the sample is given by the expression \cite{gradient}:
\begin{equation}
F(t)=\int _{V_{s}} M_z(\bm{r},t) \frac{\partial
B_{cyl} }{\partial z} dV. \label{force}
\end{equation}
Here $M_z$ is the $z$ component of the bulk magnetization and $V_s$ is
the volume of the sample. For small sample size, we make the
approximation that the field gradient $g= \partial B_{cyl} /\partial
z$ is uniform over $V_s$. A new length variable $\zeta =
B_{0}(\bm{r})/g$ is defined so that a plane of constant $\zeta$ maps
onto a surface (actually a paraboloid) of constant polarization field
which also corresponds to a sheet where the spin have the same
motion. A paraboloid of fixed $\zeta$ value, however, shifts axially
away from the iron cylinder when $B_{ext}$ increases.  In this
experiment, the sample is placed 0.70mm above the iron cylinder and
centered on the cylinder axis. At this distance, the calculated axial
field gradient is \( g = -\)470T/m (see appendix \ref{appA}).

The mechanical force detection is obtained by measuring the elastic
deformation along the $z$ axis of a micro-fabricated cantilever on
which the sample is attached. In this orientation, the probe is
sensitive to the longitudinal component of the nuclear magnetization
in contrast with a standard coil detection. The cantilever equation of
motion is represented by a damped harmonic oscillator with a single
degree of freedom. The measurement technique uses the optical
deflection of a 4 \( \mu \)W HeNe laser beam which reflects off the
rear side of the cantilever onto a position-sensitive detector.

\begin{figure}
\includegraphics*[scale=0.6,bb= 0 0 400 400,draft=false,clip=true]{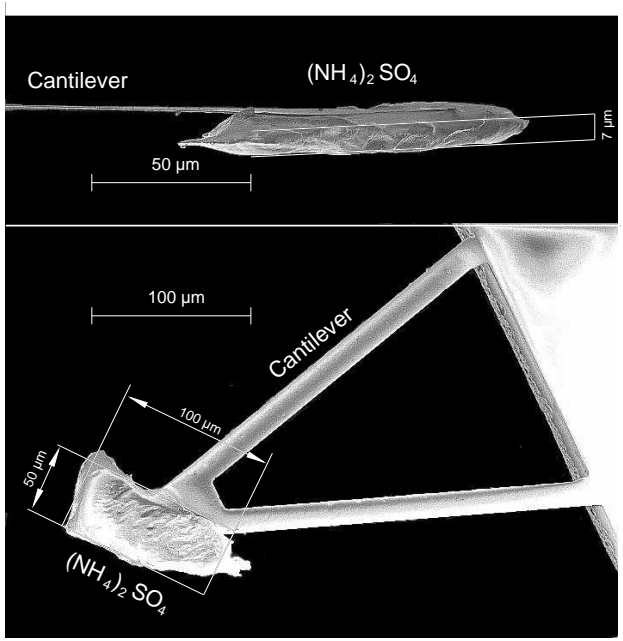}
\caption{Images of a commercial Si\( _{3} \)N\( _{4} \)(amorphous) cantilever: A 7\( \mu  \)m thick single-crystal of (NH\( _{4} \))\( _{2} \)SO\( _{4} \) sample is glued on the cantilever  end with epoxy. The loaded cantilever has a resonance frequency of 1.4kHz, a spring constant of 0.008N/m and a quality factor of 4000 in vacuum. The estimated sample volume is \(3.5 \times 10^{-8}\)cm$^3$}
\label{FIG2}
\end{figure}

Our test compound is (NH\( _{4} \))\( _{2} \)SO\( _{4} \). This
non-magnetic insulator has a high proton density \( d = 6.4\times
10^{22} \) \( ^{1} \)H/cm\( ^{3} \) and is in its paraelectric state
above 223K. NMR measurements of the $^1$H spin-lattice relaxation time
at 300K in our powder \cite{aldrich} give $T_{1z} \approx 5$s along
the static field \cite{dooglav}. The $^1$H linewidth is 5G and the
second moment is $M_2=4$G$^2$ at 295K \cite{richards60}. Our sample is
a crystal cleaved to a platelet aspect ratio and glued with epoxy on
the end of a commercial Si\( _{3} \)N\( _{4} \) amorphous cantilever
of spring constant \( k= \)0.008N/m as can been seen in
Fig.\ref{FIG2}.  After completing the assembly, the cantilever
resonance frequency drops from 5.8kHz to 1.4kHz due to the sample mass
\cite{stability}.  The quality factor of the loaded cantilever is
$Q\approx$4000 in vacuum.  From the electron microscopy images
(Fig.\ref{FIG2}), the sample dimensions are approximately \( 100
\times 50 \times 7 \mu \)m$^3$ with the smallest length (the
thickness) oriented along the axial field. This represents a volume \(
V_s = 3.5\times 10^{-8} \) cm\(^3\) or a mass \( m = 70 \)ng and
corresponds to $N \approx10^{15}$ protons.  The temperature of the
cantilever holder is stabilized around \(+27 ^\circ\)C during the
measurement \cite{thermal}. The nuclear magnetization at thermal
equilibrium is expressed by the Curie law \( \bm{M}_0 = (d \mu
_{n}^{2}B_{0}/k_{B}T) \, \bm{k} \), with \( \mu _{n}=1.4\times
10^{-26} \)J/T the proton magnetic moment, \( T=300 \)K and \( B_{0}
=1.3 \)T the polarization field. This gives a magnetic moment \( M_{0}
V_{s}=2.3\times 10^{-16} \)J/T.

In order to increase the sensitivity, $M_z$ is modulated at a
frequency $\omega_m$ close to $\omega_c$, the frequency of the
fundamental flexure mode of the cantilever. At the moment the optimal
configuration uses cantilevers that have mechanical resonance
frequencies in the audio range and Larmor frequencies \( \omega_0 \)
which are several orders of magnitude larger (radio or microwave
frequencies). Only two methods have been used to create an oscillatory
force on the cantilever: cyclic saturation and cyclic adiabatic
inversion. They are restricted respectively to compounds that have
spin-lattice relaxation times $T_1$ either shorter or larger than the
oscillation period of the cantilever.

In our case, the modulation of \( \bm{M} \) is generated by a
continuous-wave (c.w.) sequence that consists of periodic adiabatic
fast passages \cite{abragam}. The radio-frequency (r.f.)  source is a
35-75MHz Voltage Controlled Oscillator (VCO). The r.f. output field is
amplified up to 7W and fed into an impedance matched resonating
circuit (\( Q_{\mbox{rf}} \simeq 100\)) tuned to a fixed frequency,
54.7MHz. A small coil (3 turns, 0.8mm in diameter) is in series with
the tank circuit. The sample is 0.5mm away from this antenna. The
nuclear spin are irradiated for a few seconds by a linearly polarized
r.f. field \( B_x = 2 B_{1}\cos \int_0^t \, \omega(t') dt' \)
with \( \omega(t) = \Omega \sin (\omega_{m}t)+\omega_0 \), a sine-wave
modulation of the r.f.  frequency around the proton Larmor frequency
\( \omega_0 = \gamma \, g \, \zeta_0 \), where \( \gamma/2\pi =4.258
\)kHz/G is the nuclear gyromagnetic ratio. The surface of constant
$\zeta=\zeta_0$ is called the resonant sheet. The sinusoidal frequency
modulation is started at a time $t=0$. In a transformation to a rotating
coordinate system with an instantaneous angular velocity \( \omega(t)
\bm{k} \), the apparent magnetic field is:
\begin{equation}
\bm{B}_{e}(\zeta,t)= B_{1}\bm{i}+\left\{
  g \,\zeta-\frac{\omega(t)}{\gamma } \right\} \bm{k}. \label{heff}
\end{equation}
$\theta$ is defined as the polar angle made by the apparent field with
the external field. The magnetization, however, precesses about the
direction $\bm{B}_{e}+ \dot{\theta}/\gamma \,\bm{j}$, with
$\dot{\theta} = \partial \theta/\partial t$ (see appendix \ref{appB}).
A parameter for non-adiabaticity is defined with $\tan \alpha =
\dot{\theta} / (\gamma \mid \bm{B}_{e} \mid )$ the angle between the
two vectors. Provided that the adiabatic condition $\alpha \ll 1$ is
satisfied, the spin system remains at all times in a state of internal
equilibrium and \( \bm{M}\) is parallel to \(\bm{B}_e\) as required by
Curie's law. The longitudinal magnetization is $M_{z}(\zeta,t) = \mid
\bm{M} \mid \cos \theta$, where
\begin{equation}
 \cos \theta = \frac{ g \,\zeta - \omega(t)/\gamma}{ \sqrt{
 \left\{ g \,\zeta - \omega(t)/\gamma \right\}^{2}+B^{2}_{1}}}.
 \label{cos}
\end{equation}
For free spin, $\mid \bm{M} \mid$ is a constant of the motion
\cite{abragam}. This is no longer true in condensed matter because of
spin-lattice relaxation. In our sample, however, the magnetization
decay is slow compared to the modulation period. Under our measurement
protocol, an extra defocusing originates from the lack of adiabaticity
of the modulation. In a first step, these effects are neglected and
they will be considered in a more detailed analysis deferred to a
later section (see also the appendix \ref{appB}). At time $t=0$, $B_1$
is assumed to be turned on adiabatically with the sample initially in
thermal equilibrium. In this case the norm $M$ reflects the state of
the longitudinal magnetization immediately before the force
measurement.  During the c.w. sequence, the oscillatory movement of
$M_z(t)$ comes from the $\cos \theta$ factor. The value is expanded in
time series $\cos \theta \approx a_0 + a_1 \sin (\omega_m t)$
\cite{coherent} with $a_1$ the first harmonic Fourier component
\cite{zuger96} (higher harmonics have a negligible effect on the
motion of the cantilever).  Because of the large field
inhomogeneities, the amplitude of oscillation depends on the location
inside the sample.  The resonant sheet, which is the paraboloid of
constant $\zeta_0$, corresponds to the surface of maximum amplitude of
oscillation. The spatial dependence of $a_1 (\zeta)$ is the
sensitivity profile. $\Gamma$ is the half width at half maximum of
this bell-shaped curve.  $\Gamma$ has the units of a distance and it
defines the thickness of the slice probed. The amplitude of $\Gamma$
depends on both $\Omega$ and $B_1$ \cite{Gamma}.  The induced
vibration is synchronously amplified by a lock-in technique through a
single-pole low-pass linear filter of time constant \( \tau _{l} \).
For $\omega_m = \omega_c$, the lock-in signal grows exponentially (an
exact expression will be given in equation (\ref{at})) to the
asymptotic amplitude
\begin{equation}
A_0 = \frac{1}{\sqrt{2}} \,\frac{Q \, g}{k} \,\int_{V_s}  \, M_0 \,
a_1(\zeta)  \, d\zeta. \label{a0}
\end{equation}
In conclusion, the maximum amplitude of vibration achie\-ved by the
cantilever is proportional to the longitudinal magnetic moment inside
the probed slice at the beginning of the c.w. sequence. In the ideal
case of a uniform inversion of all spin inside the sample, the
asymptotic amplitude would be $A_{\text{tot}} = Q g M_0 V_s / (k
\sqrt{2})$.

%\subsection{$M_z$ measurement}

\begin{figure}
\includegraphics*[scale=0.35,bb= 0 100 700 800,draft=false,clip=true]{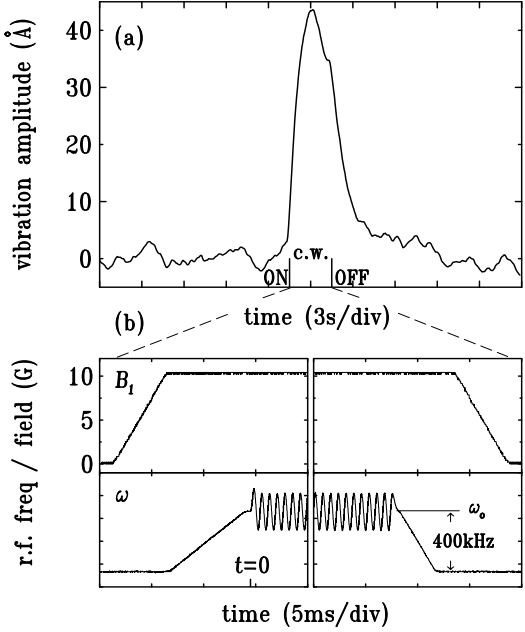}
\caption{(a) Vibration amplitude of the cantilever measured by the lock-in for a (NH\( _{4} \))\( _{2} \)SO\( _{4} \) crystal containing $10^{15}$ protons at 300K in \( B_{ext} =0.9425\)T. The trace corresponds to a single shot experiment with no averaging. The lock-in time constant is \( \tau _{l}=0.3 \)s. (b) Details of the start and end of the c.w. sequence. The crystal is irradiated for 3s by a r.f. field of $B_1 = 10$G (upper panel). The bottom panel shows the frequency waveform applied to the VCO around $\omega_0/2\pi=54.7$MHz. The amplitude of the frequency modulation is \( \Omega/2\pi =150 \)kHz. The width of the curve is the digitalization noise of the oscilloscope (not the phase noise). }
\label{FIG3}
\end{figure}

In the upper panel of Fig.\ref{FIG3} the time dependence of the
lock-in output $A(t)$ is shown when the c.w. sequence is applied.  The
time delay between force measurements is set to 27s ($> 5 T_1$) to
ensure a steady state magnetization close to the thermal equilibrium
value. The lock-in time constant is \(\tau_l= \)0.3s which corresponds
to an output noise of 4\AA. The bottom panel of Fig.\ref{FIG3}
displays the time dependence of $B_1$ and $\omega$ at the beginning
and end of the c.w. sequence. At the start, the amplitude of \( B_{1}
\) is turned on from 0 to 10G in 5ms when the frequency is well
off-resonance, \emph{i.e.} 400kHz below $\omega_0/2\pi$. The frequency
is then ramped to resonance in 7ms. Finally, the frequency modulation
of the r.f.  field is applied for 3s with a deviation \( \Omega/2\pi =
150\)kHz.  For these settings, the calculated value of $\Gamma =
7\mu$m is comparable to the sample thickness.

Since the r.f. tank circuit is tuned to a fixed frequency, the
resonance is found by sweeping the external field \( B_{ext} \).
There is \emph{no} spurious vibrations of the cantilever induced by
the r.f.  fields when \(B_{ext} \) is outside the resonance range.
Fig.\ref{FIG3}a shows the amplitude of the lock-in signal achieved in
a one shot experiment at the resonance maximum, $B_{ext} = 0.9425$T.
The maximum vibration amplitude is around 40\AA\, which corresponds to
a signal to noise ratio of 20dB.  The shape of the lock-in signal
$A(t)$ depends on the value \(B_{ext} \) \cite{wago98}.  For \(B_{ext}
\)=0.9425T, \emph{i.e.}  \(\zeta_0\) set at the middle of the sample
thickness, \emph{no} steady-state vibrations of the cantilever are
induced by the c.w. sequence and the lock-in signal decays toward zero
for long sequence. On the other hand, for \(B_{ext} \neq 0.9425\)T, an
unbalanced partial repolarization of the magnetization occurs during
each cycle and the lock-in signal decays to a finite value which
changes sign for $B_{ext}$ smaller or larger than 0.9425T.

\section{Relaxation measurements} \label{section3}

\begin{figure}
\includegraphics*[scale=0.35,bb= -40 240 620 800,draft=false,clip=true]{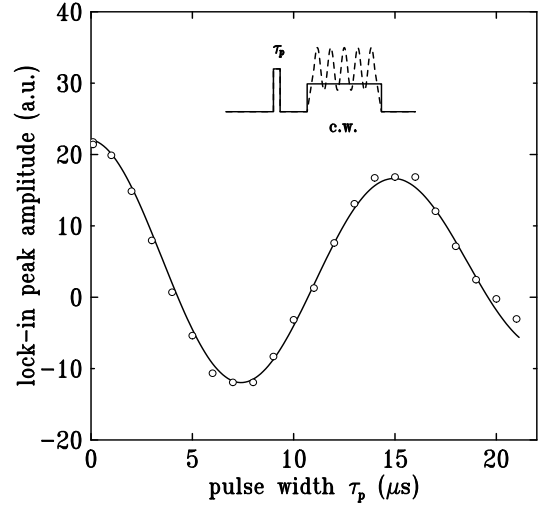}
\caption{The amplitude of the force signal (lock-in peak amplitude averaged over 1s around its maximum) is shown as a function of the width of a r.f. pulse applied 13ms before the c.w. sequence. Each point is the average of $16$ c.w. sequences. The solid line is proportional to a damped \( \cos \varphi \) with \(\varphi = \gamma \tau _{p} B_{1} \) the nutation angle. The 6.4W r.f. power during the pulse corresponds to a rotating field of $B_1=15$G at the sample location. The settings for the c.w. sequence are a r.f. field of 10G, $\Omega/2\pi = 50$kHz and $\tau_l = 100$ms. The inset is a schematic of the time dependence of $B_1$ (solid line) and $\omega$ (dashed line) }
\label{FIG4}
\end{figure}

In this section the nuclear spin dynamics of our sample is measured by
applying a series of r.f. pulses before the c.w. sequence described
above.

%\subsection{Nutation angle}
In order to calibrate the strength of the r.f. field, a r.f.  pulse of
duration \( \tau_p \) is applied, with an amplitude $B_1$, 13ms before
the c.w. sequence.  During this pulse, $\bm{M}_0$ rotates about
\(\bm{B}_e\) through an angle \( \varphi \). The angle obtained at the
end of the pulse is \( \varphi = \gamma \mid \bm{B}_e \mid \tau_p \)
\cite{wago96,wago98}. Within a few milli-seconds after the pulse, the
nutated magnetization vector decays to its longitudinal component
which then determines the amplitude of the maximum vibrations achieved
by the cantilever during the force measurement. $B_1$ is set at
maximum power, \emph{i.e.}  6.4W, for the pulse. The c.w.  sequence
uses a 2.9W r.f. field. Fig.\ref{FIG4} shows the lock-in output
averaged over a 1s time interval around its peak amplitude. For spin
that are at \( \zeta_0 \), the signal is proportional to \(\cos
\varphi \). The data are fitted by the functional form \(
\exp(-\tau_p/\tau) \cos( \gamma B_1 \tau_p) +b \).  The period gives a
calibration of the r.f. field strength at the sample location and we
get $B_1=15$G during the pulse. The other fitting parameters are \(
\tau = 43 \pm 6\mu\)s and a positive offset \( b = 3\pm 0.2 \)\AA.
The values of these last two parameters depend strongly on $B_1$.  The
positive offset $b$ is mainly due to the non-uniform field inside the
sample \cite{relaxation}. For \( ^1 \)H away from $\zeta_0$, the
direction of $\bm{B}_e$ is not exactly perpendicular to $\bm{k}$ and
only a partial inversion of the $z$ component is obtained when
$\varphi=\pi$.  The decay of the magnetic moment fitted by $\tau$ is
due to field inhomogeneity which causes a dephasing of the
magnetization in the transverse plane \cite{tau}.

\begin{figure}
\includegraphics*[scale=0.35,bb= -40 250 620 800,draft=false,clip=true]{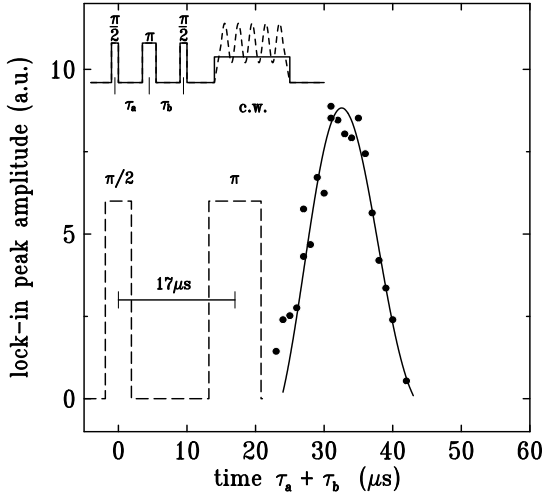}
\caption{ Measurement of the transient shape of the spin-echo: A $\pi/2$-$\tau_a$-$\pi$ pulse sequence is used to form a spin echo. The transverse magnetization is measured with the combination of a $\pi/2$ pulse and the c.w. sequence. The amplitude of the force signal is shown as a function of $\tau _{a}+\tau _{b}$ with a fixed $\tau _{a}=17\mu$s.  A r.f. field of $B_1 = 15$G is used for the pulses. The settings for the c.w. sequence  are a r.f. field of 10G, $\Omega/2\pi = 50$kHz and $\tau_l = 100$ms. The solid line is the expected shape of the spin echo for a platelet of 6.5 $\mu$m thickness.}
\label{FIG5}
\end{figure}

To study the transverse magnetization decay of $^{1}$H \cite{wago98}
a sequence of 3 pulses is used. A \( \pi/2 \) pulse is applied to the
spin system, so that the magnetization at $\zeta_0$ is rotated to the
transverse plane. After a fixed delay $\tau_a$, a \( \pi \) pulse is
applied to form a spin echo. Shortly after, a \( \pi/2 \) pulse takes
an instant snap-shot of the transverse magnetization by rotating it
along \( \bm{k}\) and the frozen component is measured with the c.w.
sequence. Varying the time delay \( \tau_b \) between the last two
pulses reconstructs the transient shape of the spin echo. Using the
same settings as the earlier measurement, the widths of the \( \pi/2
\) and \( \pi \) pulse are set to 3.8\(\mu\)s and 7.6\(\mu\)s
respectively. The delay between the center of the first two pulses is
\( \tau_a = 17 \mu\)s.  In Fig.\ref{FIG5}, the lock-in peak (again
averaged over 1s around its maximum) is shown as a function of \(
\tau_a + \tau_b \). As expected for a spin echo, the reconstructed
transverse magnetization becomes refocused at a time \( 2 \tau_a \).

%\subsection{Dipolar relaxation}

\begin{figure}
\includegraphics*[scale=0.35,bb= -40 220 620 800,draft=false,clip=true]{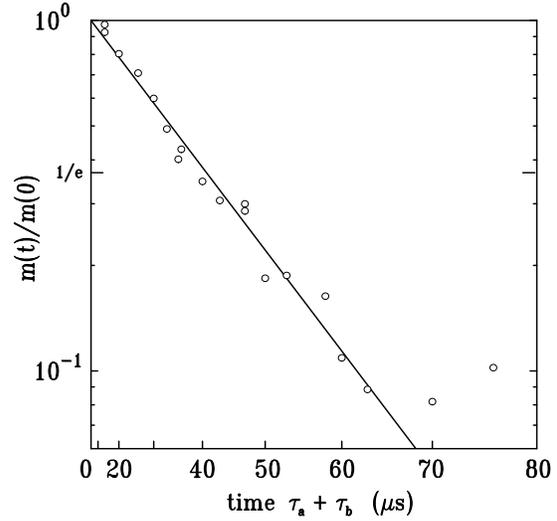}
\caption{Spin-spin relaxation time measurement: Normalized heights of the spin echo are displayed  on a square-logarithmic scale as a function of $\tau _{a}+\tau _{b}$ with $\tau_{a}=\tau _{b}\ $. The straight line is a fit with \( \exp \left\{ -(2 \tau_a/T_{2})^{2}\right\} \) where \( T_{2}= 39 \pm 1 \mu \)s.}
\label{FIG6}
\end{figure}

With increasing spacing $\tau_a$ between pulses, the size of the spin
echo signal decreases due to spin-spin relaxation. Using the same
sequence as above, Fig.\ref{FIG6} is a plot of the spin echo amplitude
measured as a function of the time $2 \tau _{a}$.  The measured values
are plotted on a \( x^2 \)- \( \log(y) \) scale and one finds that the
data follow the relationship \( \exp \left\{ -(t /T_{2})^{2}\right\}
\) with \( T_{2}= 39 \pm 1 \mu \)s. With the inferred $T_2$, the shape
of the echo in Fig.\ref{FIG5} can be calculated taking into account
the dipolar linewidth of the protons in our compound \cite{richards60}
and the spatial dependence $a_{1}(\zeta )$.  The solid line in
Fig.\ref{FIG5} is the best fit obtained for a sample thickness of $6.5
\mu$m which is in good agreement with the value obtained on the image.

\begin{figure}
\includegraphics*[scale=0.35,bb= -40 250 620 800,draft=false,clip=true]{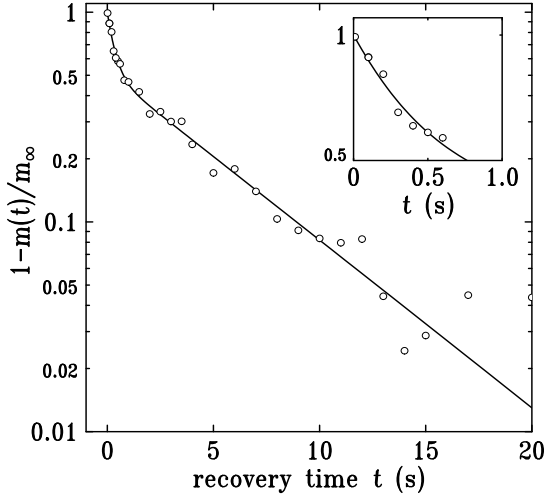}
\caption{Measurement of the longitudinal magnetization recovery: The logarithmic of the normalized amplitude of the force signal is shown as a function of the interval $t$ between  a saturation comb and the c.w. sequence. The solid line is a fit with a double exponential recovery which yields \(T _{1s} = 0.35 \pm 0.03\)s and \(T _{1l} = 5.4 \pm 0.5 \)s. Each point is the average of 32 c.w. sequences. A r.f. field of $B_1 = 15$G is used for the comb. The settings for the cyclic adiabatic inversions are a r.f. field of 7G, $\Omega/2\pi = 50$kHz and $\tau_l = 100$ms. We make sure that there is no net repolarization of $M$ during the c.w. sequence.}
\label{FIG7}
\end{figure}

%\subsection{Spin-lattice relaxation}
The longitudinal magnetization recovery is now measured after a
saturation comb \cite{fukushima}. This protocol puts efficiently
inhomogeneous spin systems in a well defined uniform state outside
thermal equilibrium. The saturation comb is composed of three \(\pi/2
\) pulses spaced by 100\(\mu\)s. The c.w. sequence is applied at a
variable delay (13ms \( < t <\) 20s) after the comb.  In order to
obtain an intrinsic measurement of the relaxation, it is important to
ensure that the sensitivity profile \( a_1 (\zeta) \) is exclusively
included inside the sample section, otherwise a partial
re-polarization of the magnetization occurs during the measurement
cycle \cite{wago98}. For our settings, \( \zeta_0 \) is set exactly at
the middle of the sample and $\Gamma = 2.4\mu$m is smaller than the
sample thickness. As before, the value plotted is the lock-in output
averaged over a 1s time interval around its maximum. No signals are
detected when $t=13$ms. On Fig.\ref{FIG7}, two relaxation times in the
recovery process are clearly observed. The results are fitted with a
double exponential \({\varrho}_s \left\{ 1- \exp(-t/T_{1s}) \right\} +
(1 -{\varrho}_s) \left\{ 1- \exp(-t/T_{1l}) \right\} \) which gives \(
{\varrho}_s = 49 \pm 2\)\%, \(T _{1s} = 0.35 \pm 0.03\)s and \(T _{1l}
= 5.4 \pm 0.5 \)s.  The value ${\varrho}_s$ does not correspond
directly to the proportion of spin that have a short relaxation
(${n}_s$) since the factor between $M_z$ and the lock-in amplitude is
also a function of $T_1$.

Neutron diffraction studies \cite{schlemper66} of the
(NH\(_{4}\))\(_{2}\)SO\(_{4}\) crystal structure show that there are
two NH$_4^+$ sites at room temperature surrounded respectively by five
and six SO$_4^{2-}$ ions. The protons of the two inequivalent ammonium
ions are coupled via dipole-dipole interactions and the measured
spin-lattice relaxation rate at 300K is an averaged value of the
$T_{1}^{-1}$. In a variable temperature NMR measurements, O'Reilly and
Tsang \cite{oreilly67} observe a single exponential $^{1}$H relaxation
process and analyze their $T_1$ results by the reorientation
correlation times $\tau_0$ of the two distinctive NH$_4^+$. At 300K,
$1/\tau_0$ should be larger than the Larmor frequency and the rotation
should be isotropic which means that $T_1$ should be independent of
the orientation between the static field and the crystallographic axis
of our sample. We suppose that our observed two $T_{1}$ processes
might be due to water contamination inside the sample during its
contact with air. The presence of H$_{2}$0 in the crystal lattice
could decrease the reorientation correlation time of the ammonium
ions, hence diminishing the protons $T_{1}$.  The relatively high
proportion of spin with short $T_{1}$ might be due to the
exceptionally small thickness of our crystal (7 $\mu$m).

In order to check the later hypothesis, a conventional NMR measurement
was performed by A. Dooglav with a 1T custom spectrometer. The sample
consisted of $\sim 1$g of our sample ground to small particles with
dimensions below $50 \mu $m. In this fine powder sample, a double
relaxation process is also observed with the following parameters
$n_{s}=17\pm 5$\%, $T_{1s}=0.37\pm 0.1$s and $T_{1l}=4.7\pm 0.2$s.
This result is in sharp contrast with experiments performed on coarser
grains, where only one relaxation is observed with $T_{1}=5.0\pm
0.2$s. The values of the two relaxation times are equal, within error
bars, to the ones measured by MRFM. In addition, measurements were
performed on the same $50\mu $m powder after two weeks of aging in
air. It showed a rise of $n_{s}$ to $26\pm 3$\% in the longitudinal
magnetization recovery experiment. A standard spin-spin relaxation
measurement on this powder seems also to indicate a double time $T_2$
with $n_{s}=20\pm 6$\%, $T_{2s}=49\pm 12\mu $s and $T_{2l}=79\pm 1\mu
$s.

\begin{figure}
\includegraphics*[scale=0.35,bb= -40 310 620 800,draft=false,clip=true]{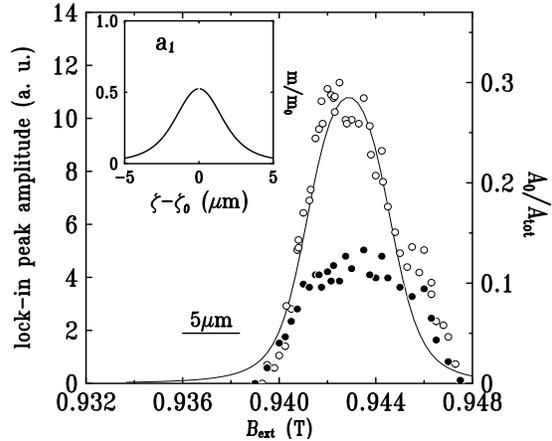}
\caption{Force signal as a function of $B_{ext}$: A saturation comb is applied 0.6s (closed circles) and 16s (open circles) before the c.w. sequence. The solid line is the expected profile for a parallelepiped sample of 7$\mu$m thickness within both the free spin and adiabatic approximations. The c.w. sequence uses a r.f. field of 7G, $\Omega/2\pi = 25$kHz and $\tau_l = 300$ms. The inset shows the transfer function that corresponds to these settings, $a_1$ is the spatial dependence of the sensitivity profile.}
\label{FIG8}
\end{figure}

One corollary issue concerns the spatial distribution of each spin
species inside the sample section. To perform this measurement, we
record the amplitude of the lock-in signal as a function of \( B_{ext}
\) for two delays $t$ between the saturating comb and the c.w.
sequence. By sweeping \( B_{ext} \), the surface $\zeta_0$ is
displaced to a different height in the sample. The force signal is
then proportional to the density of spin around this location. In
order to obtain a local measurement, the thickness $\Gamma$ of the
slice probed is reduced by decreasing both $\Omega$ and $B_1$ during
the c.w.  sequence. The inset of Fig.\ref{FIG8} shows the spatial
dependence of the transfer function $a_1(\zeta)$ for our settings
where $\Gamma $, the half width at half maximum, is $1.9\mu $m. By
changing the delay $t$, the weight $\varrho_s$ of one spin species
compared to the other can be varied. Qualitatively, the measurement
protocol gives more weight to the spin species with short relaxation
when the comb is close to the c.w. sequence. The obtained results are
shown in Fig.\ref{FIG8} for both $t=0.6$s (closed circles) and $t=16$s
(open circles).  A first look at the result indicates that a more
rounded distribution is obtained for the $t=0.6$s data. The
measurements, however, collected close to the edge of the sample are
skewed by repolarization processes that modify the shape of the
lock-in signal.  Inside the bulk of the crystal $(0.942T\lesssim
B_{ext}\lesssim 0.944T$), there is no clear evidence of a spatial
modulation of one spin population compared to the other, \emph{e.g.} a
dip of the signal in the middle of the crystal. This result suggests
that, within our resolution, the water contamination is uniform in the
thickness. The solid line is a calculation of the expected profile for
a parallelepiped sample of dimensions $100 \times 50 \times 7
\mu$m$^3$ within both the free spin and adiabatic approximations. In
spite of the idealized model, the $t=16$s data (open circles) are well
described by the calculated profile except for the high field range.
The shoulder at $B_{ext}=0.946$T corresponds to the surface of the
sample that has been glued with epoxy to the cantilever. We did not
attempt to fit this part of the data. The observed step in the signal
might be due to the protons in the epoxy. A small roll angle between
the sample and the cantilever combined with the particular shape of
our crystal is also consistent with the observed effect.

Although the data analyzed here-above ensure that two populations of
spin with different NMR properties are present, their actual
proportion is not quantitatively determined, as the actual values of
the relaxation times influence the magnitude of the measured signal.
For a better knowledge of the sensitivity of the technique to the
measurement parameters, it is then necessary to perform quantitative
analyses.

\section{Quantitative measurements} \label{section4}

%\subsection{Noise spectrum}

In this section, we shall first calibrate the mechanical response of
the cantilever and the mechanical noise. The time dependence $A(t)$ of
the lock-in signal is calculated, taking into account relaxation
processes and non-adiabatic effects.  The experimental responses for
different values of $B_1$ and $\Omega$ are compared with the
calculations.  This allows us to select an experimental condition for
which non-adiabatic effects can be neglected. It is then shown that
two relaxation times are indeed required to fit the time dependence of
the observed lock-in signals, with values consistent with those
obtained from $T_1$ data.

\begin{figure}
\includegraphics*[scale=0.35,bb= -40 200 620 800,draft=false,clip=true]{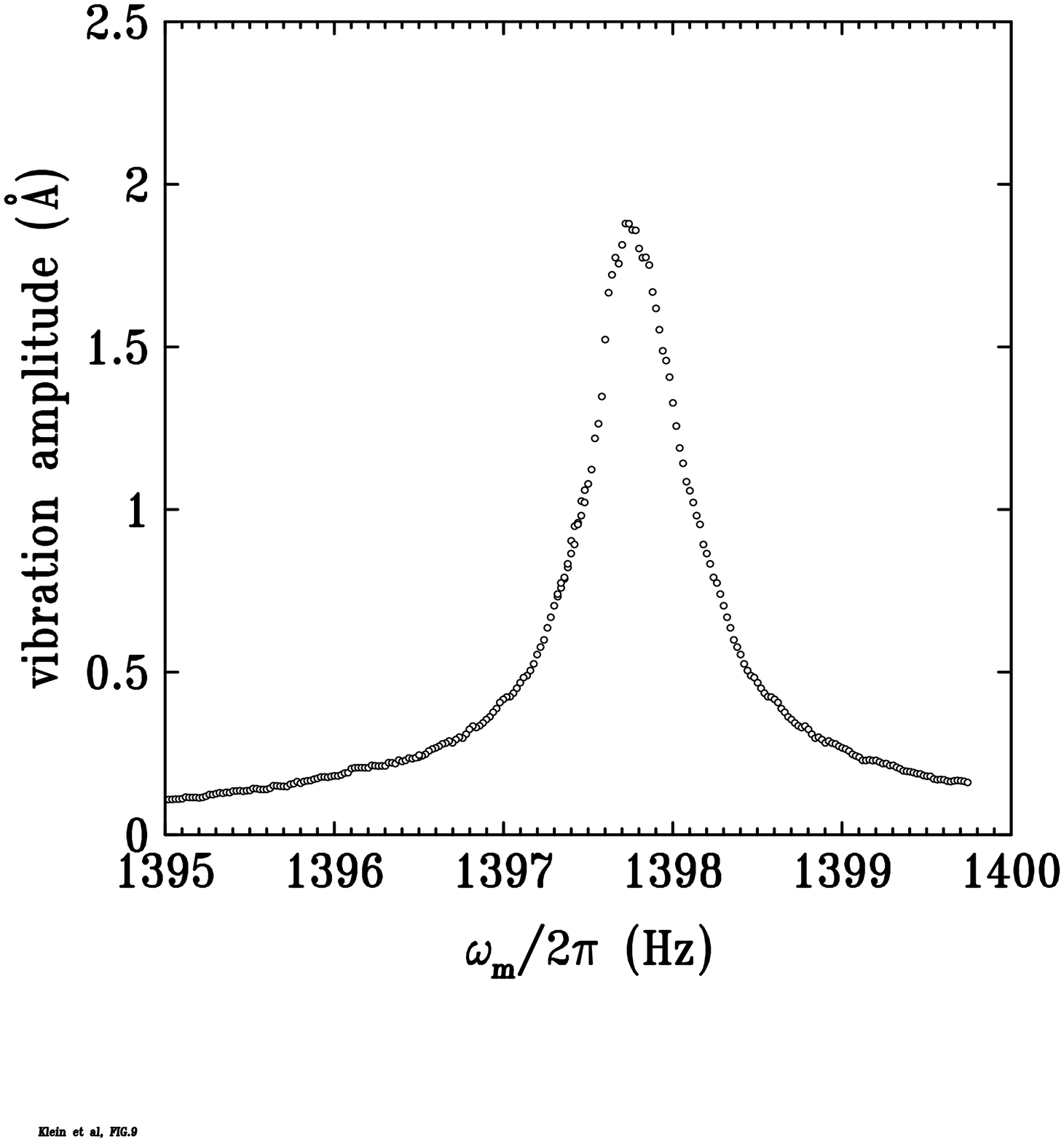}
\caption{Noise vibration spectrum of the cantilever with the sample attached: The data is obtained in vacuum (10\( ^{-2} \)torr) when \emph{no} r.f. fields are applied. The lock-in time constant is \( \tau_l  \)= 10s. The signal is the standard deviation of the lock-in output estimated over 100\( \times \tau _{l} \). The cantilever holder temperature stability is better than \( \pm 0.01 ^\circ\)C during the measurement.}
\label{FIG9}
\end{figure}

The \(Q\) of the cantilever is first measured carefully throu\-gh the
noise vibration spectrum of the cantilever loaded with the sample in
vacuum when the r.f. power is off. The lock-in time constant is set to
\( \tau _{l} \)= 10s.  An audio generator sweeps the lock-in reference
around $\omega_c$. The plotted value in Fig.\ref{FIG9} is the standard
deviation of the lock-in signal estimated over 100\( \times \tau _{l}
\) (the mean lock-in signal is zero). During the whole experiment, the
temperature stability of the cantilever is better than \( \pm 0.01
^\circ \)C which guarantees that \( \omega_{c} \) does not shift by
more than \( 0.01 \)Hz.  Fitting the squared amplitude with a
Lorentzian \cite{hutter93}, one obtains the cantilever resonance
frequency \( \omega_c/2\pi =\) 1397.77Hz and quality factor \( Q =4
000 \) (defined as the ratio of \( \omega_c \) over the full width at
half maximum of the \emph{power} spectrum).  Away from resonance, our
sensitivity is limited by the noise of the detection electronics. It
is several orders of magnitude smaller than the \AA -scale motion of
the cantilever at resonance and therefore it can be neglected. Near \(
\omega_c \), the cantilever motion consists of white noise amplified
by a narrow-bandwidth mechanical resonator \cite{sidles99}.  \( \Delta
\nu _{c} \) is the one-sided equivalent noise bandwidth (ENBW) of the
mechanical resonator \( \Delta \nu _{c}=\omega_c/8Q=0.27 \)Hz.  The
noise at the output of the lock-in is this narrow-band motion noise
observed through a \( RC \) filter of time constant \( \tau _{l}=1/RC
\) whose ENBW \( \Delta \nu _{l }=1/4\tau _{l}=0.025 \)Hz.  Exactly at
resonance, the combined distribution gives an ENBW \( \Delta \nu =
(1/\Delta \nu _{c} + 1/\Delta \nu_{l})^{-1} \)=0.023 Hz.  To convert
our data to spectral density, the resonance amplitude in
Fig.\ref{FIG9} must be divided by \( \sqrt{\Delta \nu } \). The
measured noise spectral density is \( \mathcal{A} =13 \)\AA
\(/\sqrt{\mbox {Hz}} \).  This figure also corresponds to the noise
observed in Fig.\ref{FIG3} where $\Delta\nu=0.20$Hz. The result has to
be compared with the intrinsic correlation function for fluctuations
of a Brownian particle harmonically bound to a single degree of
freedom oscillator of spring constant $k$:
\begin{equation}
 \langle A A(t) \rangle = \frac{4Qk_{B}T}{k\omega _{c}}. \label{johnson}
\end{equation}
Taking the square root of the above expression, one gets \(
\mathcal{A}_{T}=10 \)\AA \(/\sqrt{\mbox {Hz}} \) at $T=300$K which is
in good agreement with our measured value. In conclusion, our dominant
noise comes from the thermal vibration of the cantilever. From this
result, one can estimate the smallest force detectable by the
instrument in one shot \( k \mathcal{A} /Q= \) 2\( \times 10^{-15}
\)N\( /\sqrt{\mbox {Hz}} \).

%\subsection{Calculation}

\begin{figure}
\includegraphics*[scale=0.35,bb= -40 290 620 800,draft=false,clip=true]{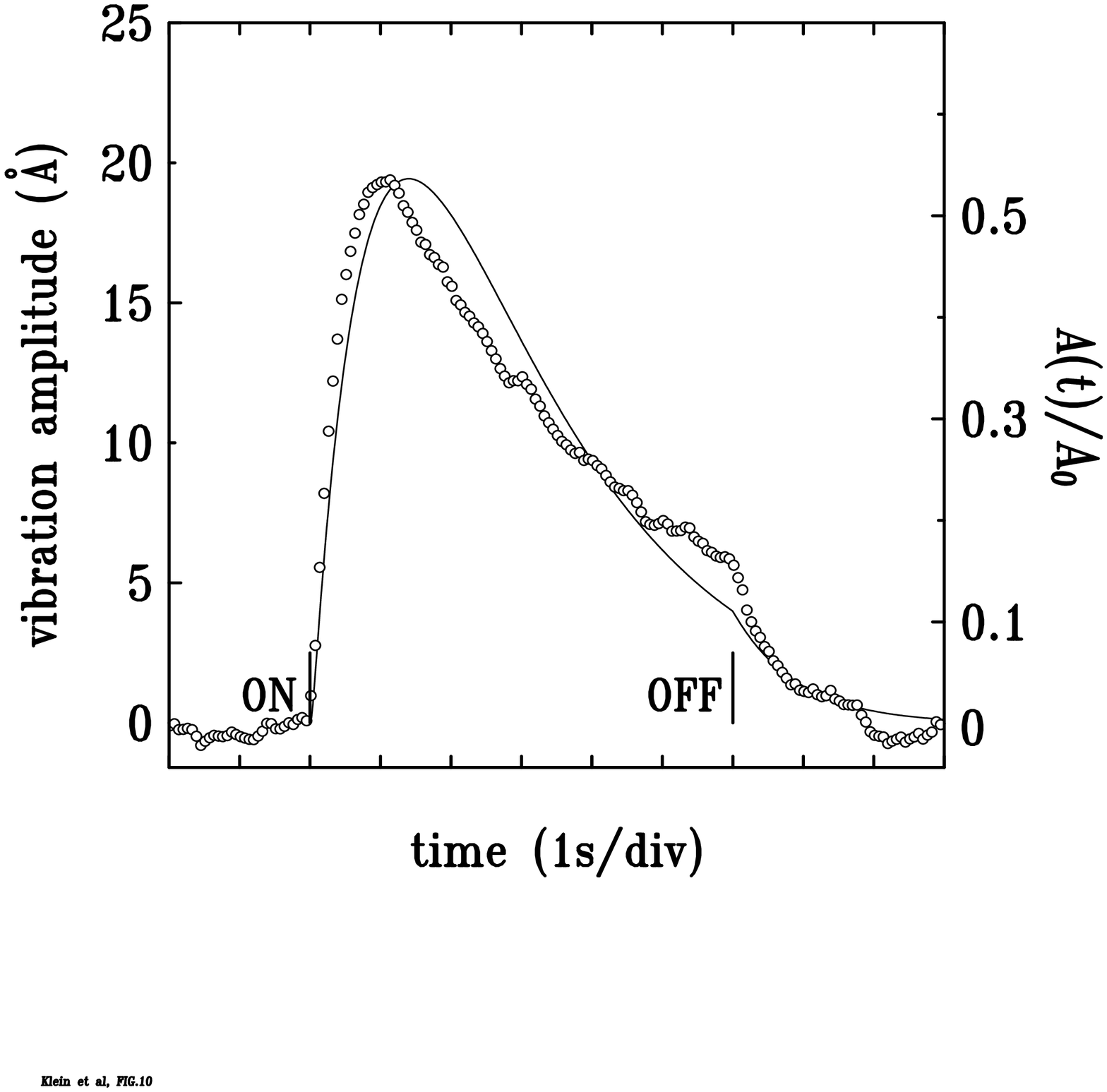}
\caption{Time dependence of the lock-in signal: The trace is the average of 32 c.w. sequences with  \( \tau _{l}=30\)ms. The c.w. sequence is composed of a r.f. field of 10G with $\Omega/2\pi = 50$kHz. The solid line is the calculated vibration amplitude of a harmonic oscillator driven by a damped sinusoidal magnetic force with a decay time constant $\tau_m=2.2\pm0.07$s. In this case, the predicted maximum signal is \(A_{\text{peak}}= 0.5 A_0 \) with $A_0 = Q F_0/k$.}
\label{FIG10}
\end{figure}

In order to obtain a quantitative measurement of our force signal when
the r.f. field is applied, a more detailed study of the time
dependence of the lock-in signal $A(t)$ is needed. The length of the
c.w.  sequence is increased to 6s compared to Fig.\ref{FIG3}.
Fig.\ref{FIG10} is the average of the lock-in signal over 32 sequences
using a short lock-in time constant of \( \tau _{l}=30 \)ms.  The
striking feature of this plot is that the norm of the magnetization \(
\mid \bm{M} \mid \) decays during the c.w.  sequence. We have checked
that experimental perturbations such as the phase noise of the r.f.
source are negligible in this case \cite{phase}.

In the rotating frame, the magnetization tends to recover slowly
towards a steady state value due to spin-lattice relaxation processes
\cite{goldman}. These relaxation processes are different from the
relaxations measured in section \ref{section3} which are linked to the
time dependence of the magnetization in the \emph{absence} of r.f.
fields. In addition, the lack of adiabaticity quantified by $\alpha
\approx \dot{\theta}/\gamma B_e$ produces a precession movement around
the locally changing effective field direction. This mistracking
corresponds to a magnetization component perpendicular to the
instantaneous precession axis $\bm{B}_{e}+ \dot{\theta}/\gamma
\,\bm{j}$ that relaxes due to spin-spin interactions. But in the limit
where $\alpha \ll 1$ and $2\pi/\omega_c \ll T_1$, the decrease of $M$
after one cycle $\Delta = M(t+2\pi/\omega_c)-M(t)$ is small. The value
is calculated at the lowest order for one spin species (see appendix
\ref{appB}).
%\begin{widetext}
\begin{eqnarray}
\Delta & \approx &
  - M  \left\{ \int_0^{\frac{2\pi}{\omega_c}} \, \dot{\theta}(t) 
\int_0^{t}  \, \dot{\theta}(t') \exp \left( - \int_{t'}^{t} \, 
\frac{1}{T_1^+} dt''  \right) \right. \nonumber \\
& & \hspace{30mm} \times \cos \left( \int_{t'}^{t} \, \frac{\gamma B_1}{\sin
\theta(t'')} dt'' \right)  dt' dt \nonumber \\
 &  &  \hspace{8mm} + \left. \int_0^{\frac{2\pi}{\omega_c}}  \, \left(
       \frac{\cos^2\theta(t)}{T_{1z}} +
       \frac{\sin^2\theta(t)}{T_{1x}} \right)dt  \right\} \nonumber \\
& &   + M_0   \int_0^{\frac{2\pi}{\omega_c}}  \,
             \frac{\cos\theta(t)}{T_{1z}} dt \label{mt},
\end{eqnarray}
%\end{widetext}
with $T_{1z}$ is the usual $T_1$ in the absence of r.f. field,
$T_{1x}$ is the transversal spin-lattice relaxation and $1/T_1^+ =
(1/T_{1y}+\cos^2 \theta/T_{1x}+\sin^2 \theta /T_{1z})/2$. It was shown
that the relaxation mechanisms in this compound are associated with
the time varying field induced by the change in the NH$_4^+$
orientation.  For an exponential correlation function with correlation
time $\tau_0$, $T_{1x}$ is expressed as a sum of the spectral density
of these fluctuating fields $J^{(i)}(\omega)=\tau_0/(1+\omega^2
\tau_0)$ with an index $i$ that corresponds to the number of net spin
flip: $1/T_{1x} = 3/2 \gamma^4 \hbar^2 I(I+1)/r^6 (5/2
J^{(1)}(\omega_0) +1/4 J^{(2)}(2 \omega_0)+1/4 J^{(0)}(2
\omega_{e}))$, with $\omega_{e} = \gamma B_e$. In the approximation
for our compound that $\omega_{e} \tau_0 \ll 1$, one obtains that
$T_{1x}$ is independent of $\theta$ and $T_{1y}=T_{1x}$.  Coming back
to equation (\ref{mt}), one recognizes that the first integral is due
to the finite value of the non-adiabaticity parameter $\alpha$ (see
appendix \ref{appB}). The second represents the spin-lattice
relaxation $T_{1\rho}$ in the rotating frame \cite{slichter}. The last
integral is the equilibrium magnetization that corresponds to the spin
temperature in the rotating frame. $\tau$ and $M_{\text{eq}}$ are
defined by rewriting the above expression in the form $\Delta/2 \pi =
- M/(\omega_c \tau) + M_{\text{eq}} /(\omega_c \tau) $. During the
c.w.  sequence, the oscillatory driving magnetic force is dampened and
the instantaneous value is given by:
\begin{equation}
F(t) \approx g \sin(\omega_m t) \int_{V_s} \, a_1 (\zeta)\, M(\zeta,t) \,
     d\zeta + \text{constant} ,
    \label{ft}
\end{equation}
with
\begin{equation}
M(\zeta,t) = \left\{ M_{\text{eq}}(\zeta)  + \left\{M_0 - M_{\text{eq}}(\zeta)
  \right\}  \exp\left( \frac{-t}{\tau(\zeta)}\right) \right\} .
\end{equation}
The integral in equation (\ref{ft}) relaxes approximately according to
a single exponential towards its equilibrium value $m_{\text{eq}}=
\int_{V_s} \, a_1(\zeta) M_{\text{eq}}(\zeta) d\zeta $ with an
apparent characteristic time $\tau_m$. One notes that
$m_{\text{eq}}=0$ by symmetry when $\zeta_0$ is centered at the middle
of the sample $\zeta_0 = \zeta_m$. The value of $m_{\text{eq}}$ is
positive for $\zeta_0 < \zeta_m$ and changes sign for $\zeta_0 >
\zeta_m$ \cite{wago98}. In the particular case where $m_{\text{eq}}=0$
and $\alpha \ll 1$, then it can be shown that the coefficient $\tau_m$
is bounded between $T_{1x} \leq \tau_m \leq T_{1z}$ \cite{verhagen99}.

The forced vibrations of an harmonic oscillator are
given by the convolution product:
\begin{equation}
 a(t) = \beta \,
        \int_0^{t} \, \frac{ F(t') }{k} \,
         {\exp\left\{ -\frac{t-t'}{\tau_c} \right\} }
        \sin\left\{ \omega_c (t - t') \right\}
      \,\omega_c  \, dt', \label{ho}
\end{equation}
with $\beta = \left\{ 1 + 1/(4 Q^2) \right\}$ and $1/\tau_c$ the
damping constant of the cantilever. In our experiment the external
force is $F(t) = F_0 \exp(- t/\tau_m) \exp( i \omega_m t)$ for
$m_{\text{eq}}=0$, with $F_0 = k A_0/Q$ and $\tau_m$ the
characteristic decay time of the magnetic force. \( {a}(s) \) the
Laplace transform of equation (\ref{ho}) is calculated in the complex
plane:
%\begin{widetext}
\begin{equation}
  \frac{k}{F_0}{a}(s) =  \frac
  { \tau_m \left ( 4 {Q}^{2} + 1 \right ) }
  {\left (s \tau_m + 1 - i \tau_m \omega_m \right )
   \left \{ (s \tau_c + 1)^2 + 4 Q^2 \right \} } \label{laplace}.
\end{equation}
%\end{widetext}
The response \( {A}(t) \) of the lock-in is the imaginary part of the
inverse Laplace transform \( \cal{L}\)\(^{ -1} \left\{ {a} (s+
  i\omega_{m}) / (1+\tau _{l}s) \right\} \) with $\tau_l$ the lock-in
time constant. An approximation can be obtained in the special case
where \( \omega_m = \omega_c \) in the limit \( Q \gg 1 \) and
\(\omega_c \gg (1/\tau_m,1/\tau_l) \).
\begin{eqnarray}
\frac{{A}(t)}{A_0} & \approx & \hspace{2.5mm} \,
\frac{1}{(1/\tau_m - 1/\tau_c) (\tau_c - \tau_l)} \exp{\left(
    -\frac{t}{\tau_c} \right)} \nonumber \\
& & + \, \frac{\tau_m/\tau_c }{(1/\tau_c - 1/\tau_l) (\tau_l - \tau_m)}
\exp{\left( -\frac{t}{\tau_l} \right)} \nonumber \\
& & + \frac{\tau_m/\tau_c }{(1/\tau_c - 1/\tau_m) (\tau_m - \tau_l)}
\exp{\left( -\frac{t}{\tau_m} \right)}. \label{at}
\end{eqnarray}
The sum of these three exponentials vanishes at $t=0$ and each term
decays to zero with a different time constant at later time $t>0$.
This leads to a peaked lock-in signal $A_{\text{peak}}$ whose
amplitude and position depends on $\tau_m$ (for a fixed $\tau_c$ and
$\tau_l$). At the end of the c.w. sequence in Fig.\ref{FIG10}
($t>6$s), the free oscillations decay of the cantilever (time constant
$\tau_c$) are observed. If one tries to fit the data with the above
nonlinear form, $\tau_m=2.2\pm0.07$s is obtained but the quality of
the fit is not very good. Values of $\tau_m$ smaller than
$T_{1x}=3.2$s have also been reported by Verhagen \emph{et al.}
\cite{verhagen99} and these findings were attributed to the phase
noise of the r.f. source. However, when large modulation amplitude are
employed for the c.w. sequence, such a fast force decay can also be
consistent with a magnetization decrease due to a lack of
adiabaticity.

\begin{figure}
  \includegraphics*[scale=0.35,bb= -40 115 620
  800,draft=false,clip=true]{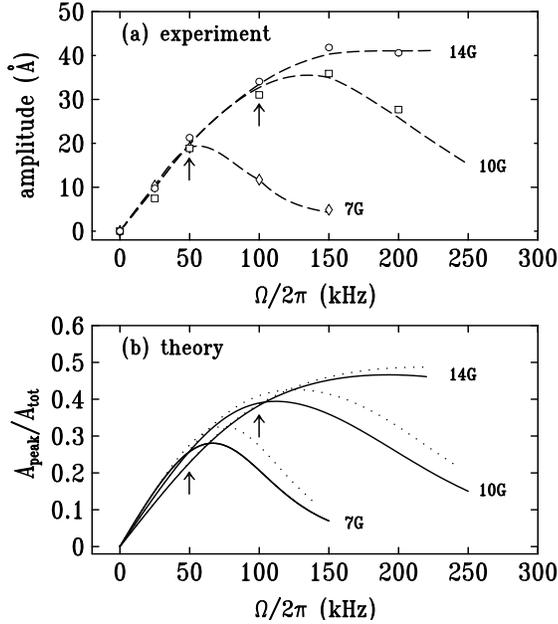}
\caption{(a) Measurements of the amplitude of the lock-in signal is shown as a function of the modulation amplitude for different strength of the r.f. field. The dashed lines are guides for the eye. The arrows indicate the limit of the adiabatic regime. The measurements are performed at $B_{ext}=0.9425$T. (b) Calculated amplitude of the lock-in signal obtained for a 7$\mu$m thick sample using equation (\ref{ft}). The parameters introduced in the model are $T_{1z}=4.9$s, $T_{1x}=T_{1y}=3.2$s and $T_{2}=40\mu$s (solid line) or $100\mu$s (dotted line).}
\label{FIG11}
\end{figure}

To understand further the meaning of this fit parameter $\tau_m$, we
plot in Fig.\ref{FIG11}a the lock-in peak amplitude measured for
different values of \( \Omega \) and \( B_1 \) when \( B_{ext}
=0.9425\)T. The value of the non-adiabaticity parameter $\alpha$
increases along the abscissa axis. For a fixed \( \Omega \) and \( B_1
\), the value of $\alpha$ oscillates with time and passes through a
maximum, $\alpha_{\text{max}} = \Omega \omega_m / \gamma^2 B_1^2$, at
time a $t=0$ modulo $\pi/\omega_m$. Fig.\ref{FIG11}b shows the
amplitude of the peak signal $A_{\text{peak}}$ predicted by equation
(\ref{at}) with $F(t)$ calculated from equation (\ref{ft}) using a
sample thickness of 7$\mu$m. The results are normalized by $
A_{\text{tot}} = Q g M_0 V_s / (k \sqrt{2})$ the amplitude associated
with a uniform inversion of all spin inside the sample. The parameters
introduced in the model are the values of the spin-lattice relaxation
times measured on powder samples by conventional NMR with
$T_{1z}=4.9$s along the static field and $T_{1x}=3.2$s along a 10G
r.f. field, in the approximation that $T_{1y}=T_{1x}$. In our
theoretical model, perturbation effects from the dipolar broadening
are also introduced. They are approximated as a time independent local
field of Lorentzian lineshape.  The solid lines are the shape
calculated with $T_2=40\mu$s (the dotted lines correspond to
$T_2=100\mu$s).

The increase of the force signal at small \(\Omega (\leq 50\)kHz)
corresponds to an increase of the modulated magnetic moment. A larger
frequency deviation increases the width of the probed slice, $\Gamma$,
and more protons oscillate at the frequency $\omega_m$. For both large
$B_1 \geq 14$G and large $\Omega \geq 150$kHz the amplitude of the
signal eventually saturates when $\Gamma$ becomes larger than the
sample thickness.

For $B_1=7$ and 10G, the deviation from the low \(\Omega\)-linear
increase (indicated by the arrows) marks the cross-over from an
adiabatic regime to a quasi-adiabatic one \cite{goldman}. In our
sample the threshold occurs at $\alpha_{\text{max}}=0.1$, in good
agreement with the theoretical model. In the adiabatic regime
($\alpha_{\text{max}}<0.1$) the value of $\tau_m$ is determined by
$T_{1\rho}$ effects, while in the quasi-adiabatic regime
($0.1<\alpha_{\text{max}}<1$) the magnetization decay $\tau_m$ becomes
smaller than $T_{1x}$. The predicted position of this cross-over
depends somewhat on the shape of the proton linewidth (dotted lines).

From the last discussion, one can conclude that the settings of
Fig.\ref{FIG10} correspond to a non-adiabatic parameter
$\alpha_{\text{max}} = 0.04$ well inside the adiabatic regime for our
compound. It can then be inferred that the decrease of the force
signal in Fig.\ref{FIG10} is due to spin-lattice effects in the
rotating frame and our fitted value of $\tau_m$ must be an average of
the two $T_1$ reported in Fig.\ref{FIG7}.

%\subsection{Lock-In signal}

\begin{figure}
\includegraphics*[scale=0.35,bb= -40 290 620 800,draft=false,clip=true]{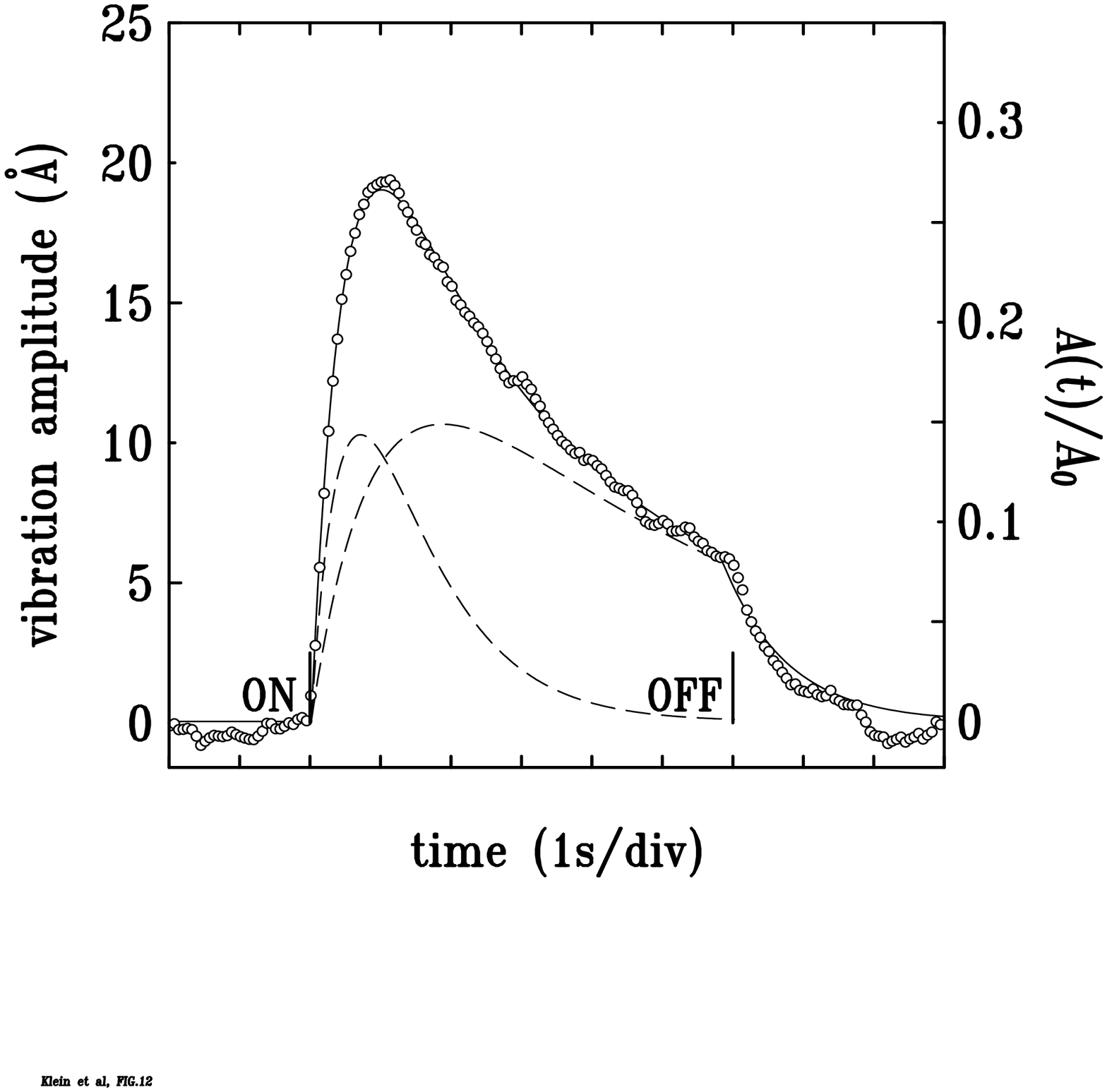}
\caption{Signal of Fig.\ref{FIG10}: The solid line is the response of an harmonic oscillator driven by two synchronous forces with short and long relaxations times $n_s A(t,\tau_{ms})+ (1-n_s) A(t,\tau_{ml})$. The best fit is obtained for $\tau_{ms}=0.55\pm0.02$s, $\tau_{ml}=4.8\pm0.2$s and $n_s = 70$\%. The height of the peak is \( 0.3 A_0\). The dashed lines show the contribution of each spin species.}
\label{FIG12}
\end{figure}

The time dependence of the lock-in signal is fitted with a double
damped synchronous excitation of \emph{two} spin populations with
respectively short ($\tau_{ms}$) and long ($\tau_{ml}$) relaxation
times. The nonlinear function $n_s A(t,\tau_{ms})+ (1-n_s)
A(t,\tau_{ml})$ is used where $A(t)$ is given by equation (\ref{at}).
The best fit is obtained for $\tau_{ms}=0.55\pm0.02$s,
$\tau_{ml}=4.8\pm0.2$s and $n_s = 69 \pm 1$\%. The result is the solid
line shown in Fig.\ref{FIG12}. The fit values obtained for the
relaxation times are similar to those measured in the magnetization
recovery experiment in Fig.\ref{FIG7}. On Fig.\ref{FIG12} the separate
contribution to the lock-in signal of each spin species (dashed lines)
are shown. The maximum force signal of the two spin species occurs
respectively 0.7s and 1.9s after the start of the c.w. sequence for
the short and long $\tau_m$. It brings up the question on how to
define the lock-in peak amplitude in the case of a sample containing
two spin species. We recall that our definition of the force signal is
the average of the lock-in peak amplitude over a 1s time interval
around its maximum value. This approach gives approximately equal
weights to both spin species in the measurement. One can also observe
in Fig.\ref{FIG12} that the height of the two peaks are approximately
equal despite the fact that there is $2.3$ times more spin with short
relaxation.  As a matter of fact it can be shown that the mechanical
detection is $2.4$ times more sensitive to spin that have a $4.8$s
relaxation time compared to spin that have a 0.55s one. From this
result, the value of the fit parameter ${\varrho}_s \approx 0.5$ in
Fig.\ref{FIG7} can be converted into the proportion of spin that have
a short relaxation $n_s = 2.4{\varrho}_s/\{(2.4-1){\varrho}_s+1\} =
70$ \%; a value that agrees well with the fit $n_s$ in
Fig.\ref{FIG12}.

Finally, the expected amplitude of the force signal for our sample is
calculated. Using equation (\ref{a0}), one gets a value of
$A_0=100$\AA\, for the settings of the c.w. sequence used in
Fig.\ref{FIG12} ($B_1 =10$G and $\Omega/2\pi=50$kHz). In
Fig.\ref{FIG12} the predicted maximum of the lock-in peak is $0.3 A_0=
30$\AA\, which is close to the experimentally measured value of 20\AA.
In conclusion, our measured amplitude of the peak lock-in signal is in
good agreement with the theoretical prediction if the two spin-lattice
relaxation times of the two spin species are taken into account. Other
effects such as misalignment of the sample compared to the cylinder
magnetic axis can account partially for a decrease of the signal (e.g.
a small offset of 0.1mm from the axis decreases the amplitude of the
lock-in signal by a factor of 2).

\section{Conclusion }

Measurement sequences combining fast adiabatic passages and pulses are
reported. They allow us to measure $T_1$ and $T_2$ for microscopic
samples using a mechanical detection. This has been applied to
quantitative analyses of the detected signals for a 7$\mu$m thick
sample of (NH\( _{4} \))\( _{2} \)SO\( _{4} \). The transverse
relaxation $T_2$ has been found consistent with conventional NMR
detection on a macroscopic sample.  Our sample displays however two
spin lattice relaxation times \(T_{1s}=0.4\)s and \(T_{1l}=5\)s. While
the long $T_{1l}$ corresponds to that measured for coarse powder
samples, the short $T_{1s}$ might be due to water contamination of our
$7\mu$m thick crystal during its contact with air. This contamination
is found to be uniform in the thickness of the sample. This large
difference in $T_1$ values has allowed us to study the influence of
the spin-lattice relaxation in the rotating frame on the measured time
dependence of the lock-in signal, as well as the variation of signal
intensity with increasing non-adiabaticity of the sweep sequence. A
consistent analysis of all the experimental parameters has been
proposed and will be quite useful in future quantitative
investigations of MRFM signals. Our work also raises the problem on
how to perform reliable spin lattice relaxation measurements at
different locations in the sample. Our investigation is mainly
restrained to the bulk, \emph{i.e.}  the middle of the sample. It is
known that the time dependence of the lock-in signal (and thus the
apparent relaxation times) depends strongly on the value of $B_{ext}$.
These difficulties prevented us from interpreting quantitatively our
results on the spatial distribution of the different spin densities
close to the sample surface. This issue will be best resolved by
performing a similar experiment on a hetero-layer sample of
well-characterized composition.

\begin{acknowledgement}

We are greatly indebted to A. Dooglav for his help in the
conventional NMR experiments. We also would like to thank C. Fermon,
M. Goldman, J.F. Jacquinot and G. Lampel for stimulating discussions.
This research was partly supported by the Ultimatech Program of the
CNRS.

\end{acknowledgement}

\renewcommand{\theequation}{\thesection.\arabic{equation}}%

\appendix

\section{Inhomogeneous field} \label{appA}
\setcounter{equation}{0}

Near the axis, a uniformly magnetized (\( M_s \)) cylinder of length
$l$ and diameter $\phi$ produces a field, whose component along
$\bm{k}$, \( B _{cyl} \), decays radially as
\begin{eqnarray}
 \frac{B _{cyl}(r,z)}{4 \pi M_s}  =  \left\{ b_{\frac{1}{2}}\left(
    \frac{z+l}{\phi} \right) - b_{\frac{1}{2}}\left( \frac{z}{\phi}
  \right) \right\} \nonumber \\
+ 3  \left\{ b_{\frac{5}{2}}\left( \frac{z+l}{\phi}
  \right) - b_{\frac{5}{2}}\left( \frac{z}{\phi} \right) \right\}
\frac{r^2}{\phi^2} + \mathcal{O} (r^{4}),
\end{eqnarray}
with \( b_a (z) = z \left( 1 + 4 z^2 \right)^{-a} \). The fields are
expressed in cylindrical coordinates with the origin centered on the
cylinder upper surface (see Fig.\ref{FIG1}). In our case $M_s$ is
calculated from the applied field $B_{ext}$ needed to produce a
resonance signal at the sample position, $z=0.70$mm.
$B_{cyl}(0,z=0.70) = \omega_0/\gamma - B_{ext}=0.352$T is replaced in
the above expression and one obtains $M_s \approx 1400$emu/cm$^3$ for
our iron. Using this result, the gradient $g=-470$T/m at the sample
location can be calculated.

\section{Adiabaticity} \label{appB}
\setcounter{equation}{0}

The aim of this appendix \ref{appB} is to calculate the decrease of
the magnetization due to the spin-lattice relaxation and the lack of
adiabaticity. The solution below is proposed by M. Goldman. In the
limit of strong r.f. fields (larger than the local field), one can
neglect the spin-lattice relaxation of the dipolar energy expectation
value. In the rotating frame, the time evolution of the different spin
components are \cite{goldman}:
\begin{subequations}
\begin{eqnarray}
\frac{\partial \langle I_z\rangle}{\partial t} & = &
- \gamma B_1 \langle I_y\rangle
+ \frac{\langle I_0\rangle -\langle I_z\rangle}{T_{1z}}  \\
\frac{\partial \langle I_x\rangle}{\partial t} & = &
+ \gamma B_1 \cot \theta \langle I_y \rangle
- \frac{\langle I_x\rangle}{T_{1x}}
- \frac{i}{\hbar} \langle[\mathcal{H}_{Dz}, I_x]\rangle \\
\frac{\partial \langle I_y\rangle}{\partial t} & = &
+ \gamma B_1 \left( \langle I_z\rangle - \cot \theta \langle
  I_x\rangle \right)
- \frac{\langle I_y \rangle}{T_{1y}}
- \frac{i}{\hbar} \langle[\mathcal{H}_{Dz}, I_y]\rangle, 
\end{eqnarray}
\end{subequations}
with $\langle\bm{I}\rangle = \mbox{Tr}(\bm{I} \sigma)$ the expectation
value of the magnetization, $\sigma$ the instantaneous density matrix
in the rotating frame and $\mathcal{H}_{Dz}$ the secular part of the
dipolar Hamiltonian. The commutator incorporates the local field
contribution defined through $B_L^2 = \text{Tr} (\mathcal{H}_{Dz}^2) /
\gamma^2 \text{Tr} (I_z^2)$. In our notation $\theta$ is the angle
between the directions of the static and effective field, $B_e =
B_1/\sin \theta$, with $B_1 \cot \theta$ the projection along
$\bm{k}$. Under r.f.  irradiation, a new coordinate system $\{X,Y,Z\}$
is defined through a transformation by the unitary operator $\exp(-i
\theta I_y)$, a rotation around $y$ by $\theta$. In the doubly
rotating frame the differential equations then become:
\begin{subequations} \label{rfa}
\begin{eqnarray}
\frac{\partial \langle I_Z\rangle}{\partial t} & = &
+ c \frac{\langle I_0\rangle}{T_{1z}}
-\left\{ \frac{s^2}{T_{1x}} +  \frac{c^2}{T_{1z}} \right\} \langle
I_Z\rangle \nonumber \\ 
& & + \left( \dot{\theta} - c s \left\{ \frac{1}{T_{1x}} -
 \frac{1}{T_{1z}} \right\} \right) \langle I_X \rangle \label{rf}\\
\frac{\partial \langle I_X\rangle}{\partial t} & = &
- s \frac{\langle I_0\rangle}{T_{1z}}
- \left( \dot{\theta} + c s \left\{ \frac{1}{T_{1x}} -
 \frac{1}{T_{1z}} \right\} \right) \langle I_Z\rangle  \nonumber \\
& &  + \gamma  \frac{ B_1}{s} \langle I_Y\rangle  -\left\{ \frac{c^2}{T_{1x}} +  \frac{s^2}{T_{1z}} \right\}
\langle I_X\rangle \nonumber \\ 
& & - \frac{3 c^2 -1}{2} \frac{i}{\hbar}  \langle[\mathcal{H}_{DZ}, I_X]
\rangle \\
\frac{\partial \langle I_Y\rangle}{\partial t} & = &
-\gamma  \frac{ B_1}{s} \langle I_X \rangle
- \frac{\langle I_Y \rangle}{T_{1y}}
- \frac{3 c^2 -1}{2} \frac{i}{\hbar} \langle[\mathcal{H}_{DZ},
I_Y]\rangle \label{Iy},
\end{eqnarray}
\end{subequations}
with $s=\sin\theta$ and $c=\cos\theta$. $\mathcal{H}_{DZ}$ is the
doubly truncated part of the dipolar Hamiltonian that commutes with
$I_Z$. The term $\dot{\theta}/\gamma \bm{j}$ is the inertia field due
to the transformation to a time-dependent reference axis. In the
adiabatic regime, defined by $\alpha = \dot{\theta}/(\gamma \ B_e) \ll
1$, it can be shown that $\langle I_X\rangle = \langle I_Y\rangle = 0$
and the first two terms of equation (\ref{rf}) are the expression of
the spin-lattice relaxation in the rotating frame for strong r.f.
fields \cite{abragam,slichter}. This appendix seeks to evaluate the
term proportional to $\langle I_X\rangle$ in equation (\ref{rf}) that
represents the decrease of $\langle I_Z\rangle$ due to the lack of
adiabaticity. Our investigation will be restricted to the
quasi-adiabatic regime \cite{goldman}, where non-adiabaticity
corrections might be dominant over $T_{1\rho}$ effects but $\alpha
<1$. The equations of motion are expressed in terms of the raising and
lowering operators $\langle I^+ \rangle = \langle I_X \rangle + i
\langle I_Y \rangle$ and $\langle I^- \rangle$ its complex conjugate.
We suppose that $\gamma B_e \ll 1/\tau_0$, the reorientation
correlation rate, for our compound which leads to $T_{1x}$ independent
of $\theta$ and $T_{1y} \approx T_{1x}$. As a further simplification,
spin-spin interactions are neglected and these effects will be
calculated numerically later on. As a consequence, one has
\begin{eqnarray}
\frac{\partial \langle I^+ \rangle}{\partial t} & \approx & - i \gamma
\frac{ B_1}{s} \langle I^+ \rangle - \frac{\langle I^+
\rangle}{T_1^+} \nonumber \\
& & - \left( \dot{\theta} + c s \left\{
\frac{1}{T_{1x}} - \frac{1}{T_{1z}} \right\} \right)\langle I_Z
\rangle -  s \frac{\langle I_0 \rangle}{T_{1z}},
\end{eqnarray}
with $1/T_1^+ = (1/T_{1y}+c^2/T_{1x}+s^2/T_{1z})/2$. Furthermore, the
period of the cyclic passage is much smaller than the spin-lattice
relaxation times. Hence, both $1/T_{1z}$ and $1/T_{1x}$ are negligible
compared to $\dot{\theta}$:
\begin{equation}
\frac{\partial \langle I^+ \rangle}{\partial t} \approx - i \gamma
\frac{ B_1}{\sin \theta} \langle I^+ \rangle - \frac{\langle I^+
\rangle}{T_1^+} -  \dot{\theta} \langle I_Z \rangle -  \sin \theta
\frac{\langle I_0 \rangle}{T_{1z}},
\end{equation}
which, upon integration, gives the result:
\begin{eqnarray}
\langle I^+ \rangle & = & -\int_0^t \, \left\{ \dot{\theta}(t')
\langle I_Z \rangle + \sin \theta(t') \frac{\langle I_0
\rangle}{T_{1z}} \right\} \nonumber \\ & & \times \exp \left\{ -
\int_{t'}^{t} \, \frac{1}{T_1^+} + i \frac{\gamma B_1}{\sin \theta
 (t'') }dt''  \right\}  dt',
\end{eqnarray}
assuming that \(\langle I^+ \rangle =0 \) at a time $t=0$. The
expression for $\langle I^- \rangle$ is the complex conjugate of the
above expression. For values of $t<T_{1}$, it is a good approximation
to neglect $\langle I_0 \rangle/T_{1z}$ compared to $\dot{\theta}
\langle I_Z \rangle$ in the first bracket. Since the decay of $\langle
I_Z \rangle$ is slow, it is replaced by a constant. Finally, the time
variation of the longitudinal magnetization is given by:
\begin{eqnarray}
\frac{\partial \langle I_Z \rangle}{\partial t} & \approx &
  -  \langle I_Z \rangle \dot{\theta}(t)  \int_0^{t}  \, \dot{\theta}(t')
  \exp \left( - \int_{t'}^{t} \, \frac{1}{T_1^+} dt'' \right) \nonumber \\
& & \hspace{17mm} \times \cos \left( \int_{t'}^{t} \, \frac{\gamma B_1}{\sin
\theta(t'')} dt'' \right) dt' \nonumber \\
& &  - \langle I_Z \rangle  \left(
       \frac{\cos^2\theta(t)}{T_{1z}} +
       \frac{\sin^2\theta(t)}{T_{1x}} \right)  \nonumber \\
& &   + \langle I_0 \rangle \frac{\cos\theta(t)}{T_{1z}} , \label{iz}
\end{eqnarray}
which has been used in equation (\ref{mt}) in the text. Although the
general form of our final expression contains a $T_1^+$-exponential
decay in the first integral, the decrease of the magnetization due to
the quasi-adiabatic part of the c.w. sequence is unrelated to
spin-lattice relaxation mechanisms. This is best seen in the free spin
limit ($T_1 \rightarrow \infty$), where the first term in equation
(\ref{iz}) reduces to the integral form of the ordinary differential
equation $\partial \langle\bm{I}\rangle /\partial t = \gamma
\langle\bm{I}\rangle \times \bm{B_e}$, an equation of motion that
preserves the norm of the magnetization. Our above solution expresses
the decrease of $\langle I_Z \rangle$ with the set of initial
condition $\langle I^+ \rangle=0$ at $t=0$. Without local field
effects, the free spin solution would eventually reach a finite
``steady state'' and for these large time scales, it is then important
to take into account spin-spin interactions.

These interactions are best seen in another rotating frame
$\{X',Y',Z'\}$ where the direction $Z'$ is aligned along the
instantaneous axis of precession of the magnetization. The new system
of differential equations is obtained by applying the unitary
transformation $\exp(-i \alpha I_X)$ to the equation system
(\ref{rfa}). Here second order corrections are examined and a more
detailed analysis should take care of the new inertia field
$\dot{\alpha}$. In this triply rotating coordinate system, the
relaxation mechanisms along the $X'$ and $Y'$ directions are dominated
by local field effects and these have a characteristic time of the
order of $T_2 \sim 40 \mu$s in our sample. The exact expression is
more complicated because the local field in the triply rotating frame
oscillates with time. The system of differential equations is solved
numerically in the approximation that the local field is time
independent and of Lorentzian lineshape. The result is compared with
the expression obtained in equation (\ref{mt}). The two calculated
values of $\langle I_{Z'} \rangle$ ($\approx \langle I_{Z} \rangle$
when $\alpha <1$) are close in both cases after one passage
($\pi/\omega_m$). The main difference occurs after several passages:
when local field effects are included, the amplitude $\langle I_{Z'}
\rangle$ decays toward zero and the expectation value of the norm of
the magnetization follows approximately the instantaneous value of
$\langle I_{Z'} \rangle$.

%\bibliography{/home/klein/Latex/Bib/mrfm,/home/klein/Latex/Bib/nh4so4,/home/klein/Latex/NMRFM/99/notes}

\begin{thebibliography}{10}
\bibitem{dooglav}
The value is measured in a 1T field using a coil detection (A. Dooglav). In
  this experiment, the powder is made of coarse grains.
\bibitem{stability}
The sample mass and shape are stable for the duration of the experiment.
\bibitem{thermal}
Using the static flexion of the cantilever as a temperature sensor, we estimate
  to \(+1.5 ^\circ\)C the elevation of the sample temperature during the
  r.f.-irradiation.
\bibitem{coherent}
We can do this approximation because all slices are phase coherent.
\bibitem{Gamma}
The width of the probed slice is proportional to $\sqrt{1.2 (\Omega/\gamma)^2 +
  B_1^2}/g$.
\bibitem{relaxation}
The fast relaxation $T_{1s}$ can also partially explain the positive offset
  $b$.
\end{thebibliography}

\end{document}